# Trade-off analysis of disruption-tolerant networking protocols

A white paper

**Caitlyn A. K. Singam**

# Table of Contents

<div align="center">**Executive summary and introduction**</div>

### A. Objective

The objective of this analysis was to simulate the performance of three different ad-hoc protocols for disruption-tolerant networking (DTN) – i.e. the transfer of information through a network of nodes in contexts prone to signal interruption/signal degradation – and to perform a trade-off analysis that will yield a recommendation of the best course of action (COA) for a user of a space-based network to employ.

This is important for space-based networks in particular since transmitting information over long distances – e.g. from directly from the initial node to the destination node – will result in the terminal signal being relatively weak, which is problematic in a high noise (disruption-prone) environment since it will likely result in packet degradation or loss unless signal power is increased to compensate. Given that changing signal power for each transmission is impractical when one's network is located in space, optimization of the signal route is the best means of ensuring the transmitted signal reaches its destination rapidly and with maximum fidelity.

The recommended COA, as determined at the end of the analysis, is one that optimizes performance in a fashion that minimizes error during transmission and transmission time to the greatest extent possible. This analysis takes into consideration the relative value of transmission time and transmission error for a user of space-based communication systems (for instance, a scientific mission – the most likely type of user for a space-based relay network[1] – would prioritize data integrity much higher than transmission time since even though a longer transmission time equals greater cost, a lower level of data integrity could result in the mission's scientific objective being compromised[2]) and provides a recommendation accordingly.

### B. Customer

The consumer for this analysis is the National Aeronautics and Space Administration (NASA) Space Communication and Navigation (SCaN) office, which conducts research into disruption-tolerant networking[3] and manages the communications systems used to communicate with NASA missions (the Near Earth Network, Space Network, and Deep Space Network)[4].

### C. Principal metrics & factors of interest

In order to evaluate the different design options, two different principal metrics of interest were be taken into account: percent error on receipt (a representation of packet integrity) and transmission time (a representation of transmission speed).

Percent error and transmission time were calculated based on two key input factors (distance and signal quality). The equations for these two factors, and their relationship with the metrics of interest, is discussed in further detail in Section 1 below.

At the end of the analysis, the calculated values for percent error and transmission time for each design option were combined together using a multi-attribute value function (MAVF) to yield a single quantitative value representing the relative value of each solution. The MAVF can produce values ranging from 0 (representing the worst option) to 1 (the optimal design option). The analysis objective was considered satisfied when the design option with the highest MAVF ranking was identified.

### D. Design options

Three different design options were considered: bundle protocol, the current state of the art routing protocol for DTN[5] which picks the route that strikes a balance between distance and signal quality; distance-based Djikstra, which selects a route that minimizes the distance traveled during transmission; and signal quality-based Djikstra, which selects a route with the best overall signal quality (without regard for distance traveled or any other factors). (Section 2 describes the design options in further detail.)

### E. Techniques used

To evaluate the different design options, a generic/hypothetical space-based network comprised of 10 satellites was generated to route data from an initial node to a destination node. Apart from the distance between the initial node and destination node (which was fixed), the distance and signal quality for the link between any two nodes in the network was instantiated randomly. A Monte Carlo simulation was then used to simulate the routing of 500 packets through the network (with each packet representing one 'sample', or iteration of the Monte Carlo simulation) using the three different routing protocols being evaluated. The Monte Carlo simulation was used to vary the signal quality and inter-node distance associated with each link in the network each step in order to represent variations due to environmental phenomena and orbital movement.

At the end of each iteration, the transmission time for each packet (based on the distance traveled by the packet, and given that the packet is transmitted as an electromagnetic wave traveling at the speed of light) and the packet's state (intact or damaged) was determined. The aggregate data from the entire simulation was then used to calculate the mean transmission time and percent error for the overall sample (determined based on the terminal state of each packet on receipt), both of which were in turn fed into a MAVF. The recommended design option was chosen based on which option was associated with the highest MAVF output value. Welch's t-tests were also used to confirm that the differences in metric values observed were in fact statistically significant (i.e. that there was enough of a difference in the performance of the different options for the choice to have a significant impact on overall system performance).

Additional details are provided in Section 3.1.

### F. End result and recommendation

Using the analysis and simulation techniques described above, it was determined that signal quality-based Djikstra was the best routing protocol for DTN. It outpaced bundle protocol (the current SOA protocol) in terms of both metrics of interest, and based on known metric priorities, provides the best trade-off between transmission speed and data integrity.

### 1. System description

Since the purpose of this analysis was to determine the relative performance of different disruption-tolerant networking protocols for a generic space-based use case, the analysis used a generic, hypothetical network as its system of interest.

The system (as depicted in the domain block definition diagram shown in Figure 1) was comprised of a network of 10 space-based relay satellites, located at different distances from each other and from the Earth-based ground station. The satellites (which were modeled as nodes on a graph) all had communication links with one another, as well as with an initial node (the space-based asset generating the data being transmitted) and the final node (the ground station). The distances in between nodes, as well as the signal quality values for the links between nodes, was determined randomly; the one exception to this was the distance between the transmitting node (noted as a 'deep space asset/probe' in Figure 1) and the receiving ground station, which were placed at a fixed distance from one another in order to set the scale (i.e. maximum distance) the network would operate at. The value used for the internode distance between the initial and destination nodes was the maximum distance from Earth to Titan, $1.27 \times 10^9$ kilometers, as an actual mission – the Cassini-Huygens mission – transmitted data back to Earth at that distance[6], and the distance thus is an accurate representation of the system context that would have to be handled by any space-based network attempting to implement any of the tested routing protocols on a practical basis.

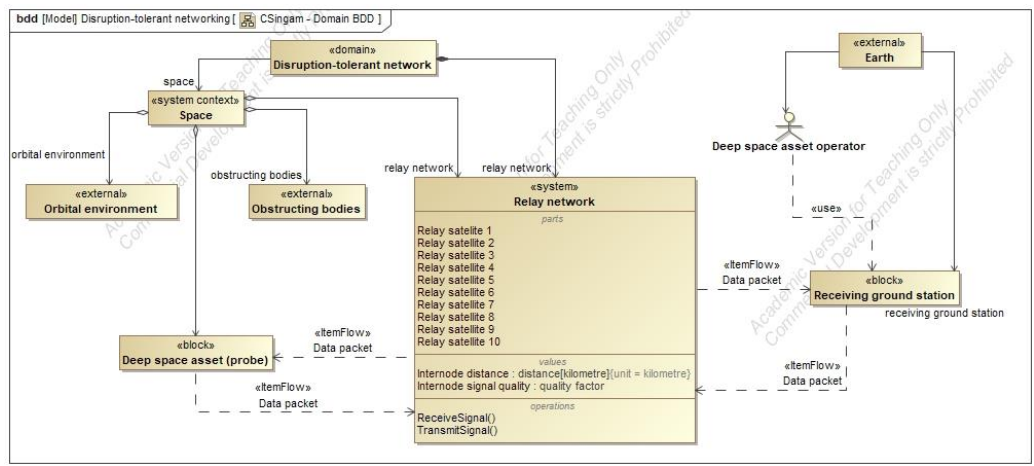

***Figure 1*** *A domain block definition diagram showing the principal components of the analyzed system.*

The signal receipt and signal transmission operations that are shown as operations of the satellites within the relay network system in Figure 1 above, when taken in aggregate for all the satellites within the network, represent the system behavior dictated by the network's routing protocol. The system's ability to transmit and receive signal among its component satellites, however, is influenced by two key input factors, internode distance (expressed in kilometers) and internode signal quality (expressed as a percentage of the theoretical ideal for signal performance).

Internode distance, determined by the user's existing asset placement, is a key factor since the time it takes to transmit a signal from point A to point B is, inherently, tied to distance – since the electromagnetic waves used to convey signals travel at the speed of light, $c$, the time taken to travel between node A and node B is dictated by the distance between the two nodes divided by $c$. (Links with shorter internode distances are thus preferable to those with longer distances, as shorter transmission time is always preferable).

Signal quality is also of interest due to the fact that it is a direct reflection of the role environmental context plays in determining the quality of a link and how much of the signal gets successfully received by the end node. It represents link quality as a percent of the theoretical ideal/maximum, ranging in value from 0% to 100, and can be calculated based on historical link performance data and information from environmental models. Though higher signal quality is always preferable, signal quality levels near the theoretical maximum of 100% are rare due to the inherent noisiness of a real-world environment. Disruption-tolerant networks need to take into account, or alternatively be resilient against, the effects of varying and/or poor signal quality lest the data payloads they are transmitting be lost or damaged.

Together, these two factors influence the two main metrics of interest for any transmission that moves through the system: percent error on receipt and transmission time.

Percent error on receipt, or the percentage of packets (out of the total number initially transmitted) which are damaged and/or lost during transmission, can range from 0% to 100%, but should be as low as possible (i.e. as close to 0% as possible) in order to respect the need for data integrity during transmission. This metric is viable for evaluating the performance of real-world networks since the number of packets (and the number of bytes per packet) in a given transmission will be defined and kept constant for a particular mission. Thus, when the users of the Earth-based ground station receive packets on behalf of a mission, they can compare the number of packets (and bytes) received with the number that should have been sent by the mission. Given that the entire point of a transmission is to ensure that the receiving node gets the information that was being transmitted, ensuring that percent error is as low as possible (and that as much of the data payload as possible is intact) is a high priority.

Transmission time, the amount of time (in hours) it takes for a single packet to travel from the initial (transmitting) node to the destination (receiving) node, has a theoretical minimum dictated by the time it would take a pulse traveling at the speed of light to travel the straight-line path from initial to terminal node. However, since relay networks route packets through non-linear paths to their destination and increase transmission time beyond the minimum, it is thus important to consider how much additional transmission time the protocol incurs (and attempt to keep the overall transmission time as close to the theoretical minimum as possible). For missions where data packets may contain time-sensitive commands, it is imperative to ensure that transmission time does not become excessively long.

Percent error and transmission time are the primary criteria by which the different routing protocols (the design options being evaluated) used by the network can be evaluated, and will be the focus of this analysis.

## 2. Design options

The three design options being evaluated as part of this analysis are bundle protocol, distance-based Djikstra, and signal-quality based Djikstra. Their relative performance with respect to the two metrics of interest (described above) is shown in Table 1 below. The exact metric values for each design option were determined via the Monte Carlo simulation, and are discussed in Section 5 of this document.

***Table 1*** *A comparison of the relative performance of the design options being evaluated, with regards to the metrics of interest.*

|  | Design options | | |
| --- | --- | --- | --- |
| **Metric** | **Bundle protocol** | **Distance-based Djikstra** | **Quality-based Djikstra** |
| **% error** | Medium | High (suboptimal) | Low (optimal) |
| **Transmission time** | Medium | Low (optimal) | Medium to high (suboptimal) |

The details of how each of the design options work, and the parameters prioritized by each, are presented below.

### 2.1. Bundle protocol

Bundle protocol is reflective of the store-and-forward methodology put forth by the Consultative Committee for Space Data Systems (CCSDS)[7] for packet routing. As shown in Figure 2, in a network following bundle protocol, packets are transmitted from the node they are the currently on to the adjacent/neighboring node that meets the criteria of being A) closer than the current node to the destination node, and B) having the highest signal quality of the nodes that meet criterion A. This process is repeated until the packet reaches its destination. This typically results in the packets being routed through more nodes than in either of the other two methods, though depending on the length of the links used, this may not necessarily correspond to a longer transmission time or worse signal quality.

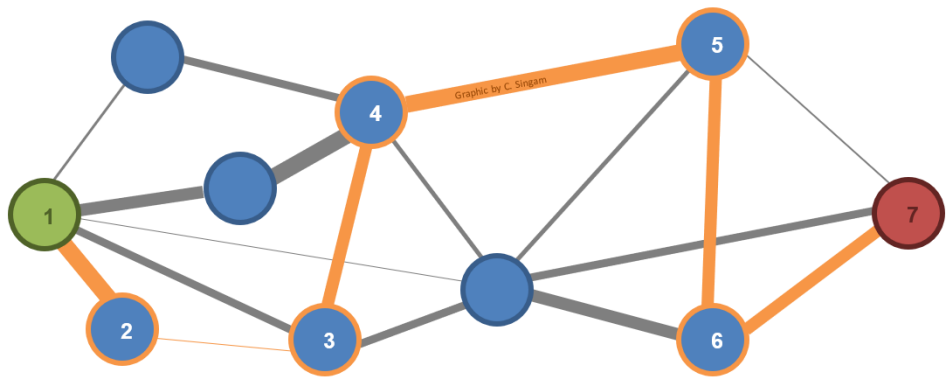

***Figure 2*** *An example of how bundle protocol routes information through a network from the initial (green) node to the terminal (red) node, with nodes numbered in order of visitation and the path marked in orange. Line widths correspond to signal quality (thicker lines correspond to higher quality links), and distances between nodes are to scale.*

### 2.2. Distance-based Djikstra

This design option uses Djikstra's algorithm to find the path with the shortest distance to the terminal node, which travels through at least one intervening (relay) node. The stipulation that the recommended path include at least one relay node is to eliminate direct-to-Earth transmission routes, which have already been shown in the literature[8] to be outperformed by relays. This is achieved by instantiating the edge costs of each link as the corresponding internode distance, with the exception of the link from the initial node to the terminal node (if present) which is instantiated such that the edge cost is significantly higher than the edge cost of any other link and thus resulting in Djikstra's algorithm rejecting the direct route as a a possible path.

The results of this algorithm (exemplified in Figure 3 below) usually yield a path that is as close to the straight-line path (marked using a dashed line in the diagram below) as possible, resulting in a short transmission time but frequently resulting in a path that includes low-quality links that degrade the signal prior to receipt.

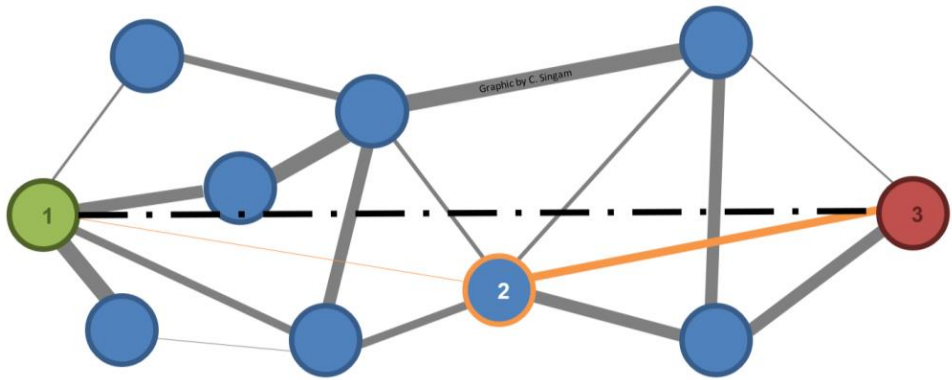

***Figure 3*** *An example of how the distance-based Djikstra method routes information through a network from the initial (green) node to the terminal (red) node, with nodes numbered in order of visitation and the path marked in orange. Line widths correspond to signal quality (thicker lines correspond to higher quality links), and distances between nodes are to scale. The dashed line marks the straight-line path from the initial node to the destination.*

### 2.3. Signal quality-based Djikstra

This design option evaluates the highest-quality path through the network using the Djikstra algorithm, using 1 – [internode signal quality as a decimal probability] (as calculated between each pair of nodes, rather than over an entire path) as edge costs for the network. Since the Djikstra algorithm selects the path with the minimum cost, and it is desired to maximize the signal quality seen on the chosen route, the edge costs are instantiated as the complement of the parameter of interest (in this case, signal quality).

Since this algorithm uses Djikstra shortest-path algorithm with complement edge costs, rather than a longest-path algorithm in conjunction with a graph that uses signal quality directly as the edge cost, the results of this algorithm favor paths (exemplified in Figure 4 below) with relatively few links and high signal quality. As a consequence, it is good at minimizing percent error but (since the algorithm favors fewer links, not necessarily shorter ones) it is suboptimal at minimizing transmission time.

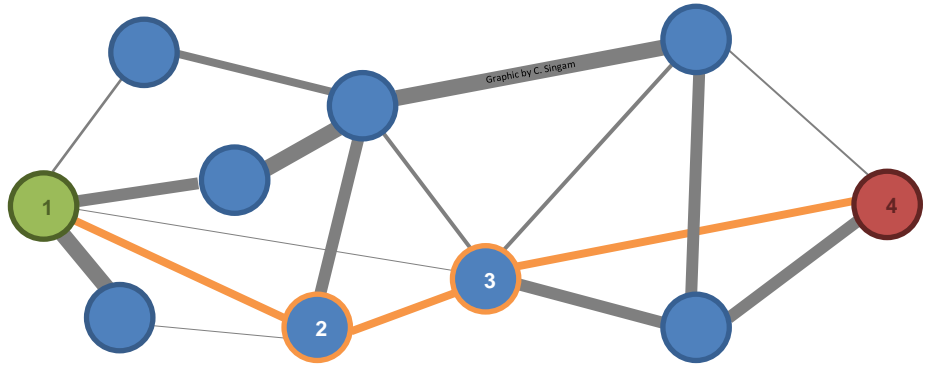

***Figure 4*** *An example of how signal quality-based Djikstra routes information through a network from the initial (green) node to the terminal (red) node, with nodes numbered in order of visitation and the path marked in orange. Line widths correspond to signal quality (thicker lines correspond to higher quality links), and distances between nodes are to scale.*

## 3. Analysis approach

### 3.1. Overview

The primary objective was to identify the mean values associated with each of the design options for the performance metrics of interest, so that a MAVF analysis could be performed and the design options could be quantitatively ranked. However, given that using signal quality-based Djikstra and distance-based Djikstra for space-based networking contexts is, to the best of the author's knowledge, a concept of the author's own devising, no data exists on the performance of such methodologies for the desired use case and in the desired system context. However, given the extravagant financial and scheduling burden that would be involved in constructing any sort of reasonable prototype of a space-based network – or otherwise acquiring even temporary access to existing networks[9] – for testing purposes, obtaining real-world data on the performance of these protocols is, disappointingly, highly impractical. Fortunately for the aerospace community, creating a model of a space-based network and simulating the movement of packets through it using various routing protocols is significantly more practical, providing the data that allows the age-old quandary (of which packet routing protocol performs best in a space-based context) at the heart of this analysis to be answered satisfactorily.

To that end, a multi-step approach was taken to develop a model of sufficient faithfulness to reality as to accurately test the mettle of the three design options of interest. Firstly, the positions of the initial and destination nodes were established in a MATLAB model; the relay network satellites were randomly instantiated so that any one satellite had an equal probability of appearing anywhere between the initial node and destination node.

After each node had been linked with each other node, and all the links had been assigned distances and random signal qualities, the movement of 500 packets through the relay network was simulated using the Monte Carlo simulation method. The packet state and transmission time were recorded for each packet when it was simulated as having reached its destination. Once all 500 packets were simulated as having traveled through the system, the percent error for the population was calculated by totaling the number of packets recorded as having been lost and dividing that by the sample size (500 packets). The mean transmission time associated with each design option was also calculated.

Since each run of the Monte Carlo simulation is associated with a different randomly instantiated graph (due to the way the simulation setup is implemented in the MATLAB code), it is possible to run the simulation code multiple times in order to gain an accurate image of how all three design options perform across several different networks. Due to the processing power and time needed to run each simulation, a relatively small sample of 5 runs was used. The values for percent error and transmission time for each run were collected in Excel, and the mean, standard deviation, and standard error of the mean for the two metrics of interest were calculated for each of the design options.

The obtained 5-run means for percent error and transmission time were subsequently fed into a MAVF for final analysis[10]. The metrics were assigned preference weights per the Parnell swing weight matrix[11]: percent error, as a Parnell 'mission critical/large effect' parameter (a metric which needs to be optimized in order to ensure mission success and which mission success is sensitive to) was assigned a weight of 100; transmission time, as a 'mission effectiveness/small effect' parameter (a metric which can be used to compare the relative worth of design options, but which mission success is not as sensitive to) was assigned a weight of 20. The results of the MAVF for each design option were then directly compared and used to make the final recommendation. Additionally, as a confirmatory measure as to whether the designs performed significantly differently from one another, the 5-run data means and standard deviations were used to perform one-tailed Aspin-Welch t-tests[12] to identify whether or not the differences seen between design options for both transmission time and percent error were statistically significant ($p<0.05$). The equation for the t-statistic comparing the means of samples from two different populations (population 1 and population 2) is:

$$ t = \frac{\overline{x_1} - \overline{x_2}}{\sqrt{\dfrac{s_1^2}{n_1} + \dfrac{s_2^2}{n_2}}} $$

where $\bar{x}$ is the mean for a sample, s is the sample's standard deviation, and n is the sample size.

### 3.2. Response model

The workflow and methodology for the trade-off analysis, as described in Section 3.3.1 above, is summarized in the following combined response model (Figure 5, below).

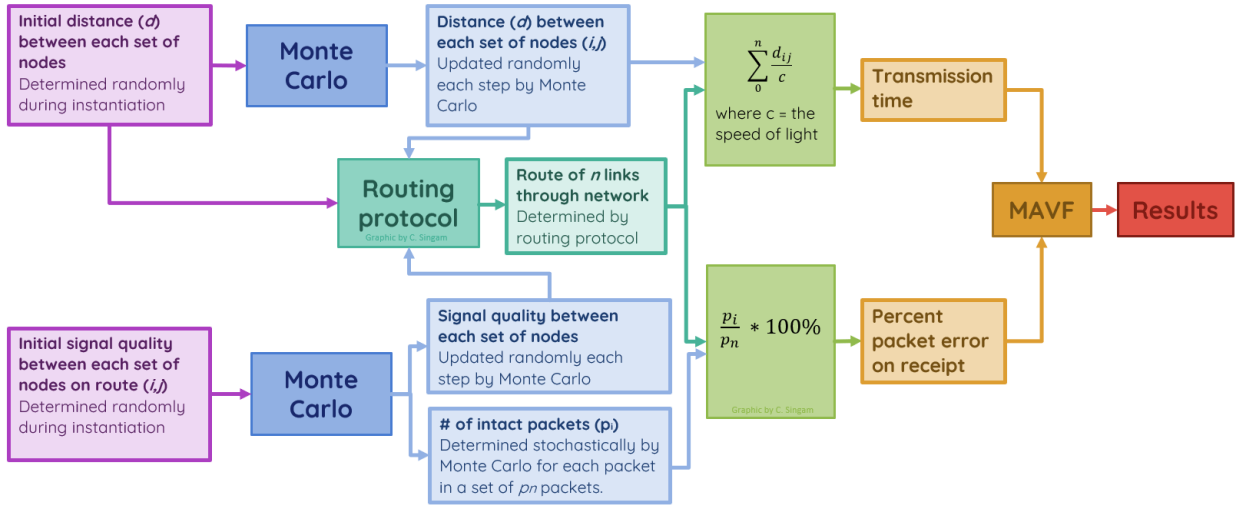

***Figure 5** A combined response model diagram for the entire analysis, showing the initial input factors (purple) and intermediate factors (blue) produced by the Monte Carlo simulation and fed into the routing protocols being evaluated (turquoise). The equations used to calculate the metrics of interest (green) as they are fed into the MAVF (yellow) are also shown.*

### 3.3. Modeling and simulation needs

As addressed in Section 3.3.1 and shown in modeling and simulation played a major role in providing the data needed to perform the MAVF analysis central to this trade-off. In order to meet the modeling/simulation and analysis needs required for the proposed project, the following models and simulations were used:

- *Network graph model:* a graph was used to represent the nodes in the network. As discussed in Section 2, two of the three design options relied on the edge costs of the graph to be instantiated with respect to a factor of interest in order to apply identify the Djikstra shortest path. Since one design option required the edge costs to be based on distance and one required the edge costs to be based on signal quality, the MATLAB implementation associated with this model maintained both a distance-based graph and a signal quality-based graph.

- *Monte Carlo simulation*: an event-step Monte Carlo simulation was used to simulate the movement of packets through the network via different routing protocols, add random variability to the distances and signal quality between nodes in the network, and to stochastically determine the state of each packet (intact or damaged) after it had been sent to a new node.

Further details regarding the inputs/outputs of the models and simulations mentioned above are addressed in Section 4.

### 4. Supporting models and simulations

As addressed above, the analysis presented herein utilized two network graph models and a Monte Carlo simulation.

### 4.1. Network graph models

As described in Section 3.3.1, the relay network that the packets are routed through is represented in the programmatic implementation as a digraph, which is compatible with MATLAB's Djikstra implementation, its shortestpath function[13]. The use of network graphs to model satellite networks is not unusual[14][15]; however, such graphs are typically used to model control capabilities rather than routing.

The two network digraphs used in this analysis are nearly identical to each other for any given run of the MATLAB program. Each accepts the list of nodes (the 10 relay satellites, the probe, and the ground station) as inputs, as well as arrays representing the links in between nodes (the same for both). The output of these both these models are the fully instantiated and functional digraphs that the Monte Carlo simulation can route its packets through and that the Djikstra algorithm can be applied to.

To prevent the direct-to-Earth link (i.e. straight line path from initial to destination node) from being used (as discussed previously), the edge cost for the link associated with the direct-to-Earth route on each graph is assigned a value that far exceed the maximum possible value for the edge costs on the other links.

The only difference between the two digraphs is the parameter used as the edge cost for each.

<u>Distance-based graph</u>

Before the graph models are instantiated, each node is assigned an x and y coordinate on the Cartesian coordinate plane (with units in kilometers). Though the coordinates for the initial node and the destination node are fixed so that they are diametrically apart on the coordinate plane and at Titan distance from one another, the relay network's satellites are all assigned coordinates randomly, per a uniform distribution. The lower limit on coordinate distances is defined as 10^4 kilometers (the same as low Earth orbit[16]), and the upper limit is defined by the location of the transmitting probe.

The distance-based digraph uses the randomly assigned position of each node to calculate internode distances and assign edge costs for each link (with the exception of the direct-to-Earth link, as described above). The distance formula is:

$$Distance_{AB} = \sqrt{(x_B - x_A)^2 + (y_B - y_A)^2}$$

For convenience in implementation, since the metric of interest is transmission time (not transmission distance) – which is just distance divided by a constant (the speed of light) – the edge costs for the distance-based digraph are actually assigned transmission times as edge costs. This does not result in any practical difference in how the model functions or how the Monte Carlo simulation interacts with it; it simply makes programmatic implementation easier.

<u>Signal quality-based graph</u>

The edge costs for the signal quality-based graph are much simpler to instantiate.

Since signal quality can theoretically vary between 0 and 1 (when expressed as decimals instead of as percentages), but must to skew slightly more to the right than a normal distribution (since satellites are designed with antennas, etc. that are optimized for performance in their environments), a beta distribution was used to randomly determine the signal quality to each link in the model. Figure 6 shows the beta distribution with the parameters selected for this analysis (a=3, b=2).

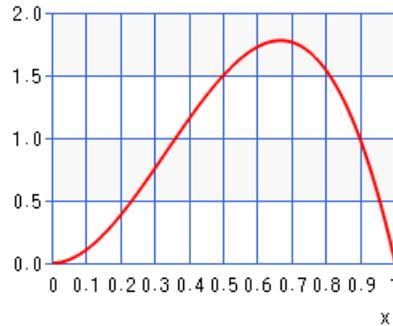

*Figure 6* *Beta distribution, a=3, b=2. Graph generated using Casio's Keisan online calculator*[17]

The complement of the signal quality values produced by the beta distribution are then taken using the equation 1-[signal quality] and assigned to the corresponding links as edge costs.

### 4.2. Monte Carlo simulation

The Monte Carlo simulation used in this analysis follows standard implementation methodology presented in the literature[18]. The inputs to the simulation are the fully instantiated digraphs described above, with edge costs reflecting the randomly assigned internode distances and signal qualities.

Each iteration of the simulation, a packet is routed through the network (as represented by the digraphs). With each step, the packet advances from the node it is currently on to the next node in the path dictated by each routing protocol. (In effect, there are three 'copies' of the packet moving through the network simultaneously – each subject to a different protocol, but all experiencing the same conditions). The each 'hop' between nodes is associated with transmission time – added to a running total that is recorded when the packet reaches its destination – and the possibility that the packet has been lost/damaged mid-transmission, which is determined stochastically by comparing a randomly generated decimal (between 0 and 1) with the assigned signal quality (not to be confused with the edge cost) of the link that has been traveled over. If the random decimal is lower than the link quality, it is deemed to have survived the 'hop', else it is recorded as having been irreparably damaged/lost. For the sake of data collection (and avoiding the complexities associated with censored data[19]), 'lost' packets are not removed from the transmission queue and are simulated like intact packets all the way to their destination.

The edge costs are updated each step to represent environmental fluctuations; they are all randomly assigned new values per a normal distribution centered around their initial (start of the simulation) or 'default' value. The digraphs are re-established accordingly with the updated edge costs, and each of the the routing protocols re-evaluate their recommended paths based on the node their 'copy' of the packet is currently on and the new edge costs.

Once all three 'copies' of the packet reach their destination, the transmission time (calculated based on distance traveled and independent of simulation time) and packet state for each copy is recorded. The digraphs are then reset to their state at the start of the simulation after each iteration for the sake of consistency between samples/packets, and the iteration ends. Each run of the Monte Carlo simulation simulates the movement of 500 packets through the system (i.e. there are 500 samples/iterations). This was determined to be a sufficient number of samples based on a cumulative running mean (CRM) plot for transmission time for all three routing protocols being evaluated: the mean for bundle protocol, the most variable of the three, stabilized after around 250 samples, which means that 500 samples provides a wide contingency margin.

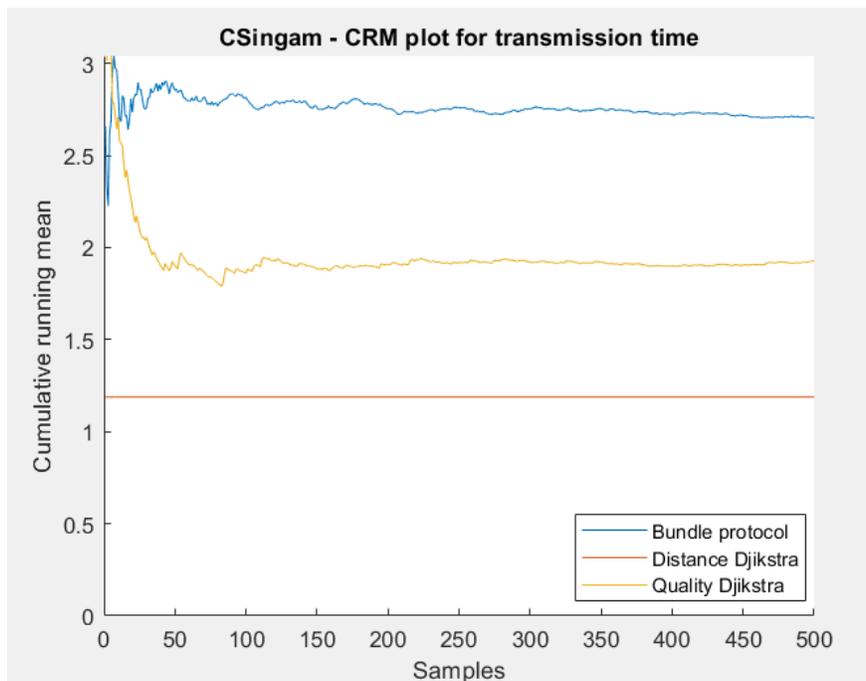

***Figure 7*** *Cumulative running mean plot for the Monte Carlo simulation, showing stabilization of the mean for all three protocols after around 250 samples.*

## 5. Analysis and results

The results for a single run of the Monte Carlo simulation are summarized below in Table 1. As seen from the results in the table, bundle protocol performed moderately well in terms of percent error but was the worst in terms of transmission time, while distance-based Djikstra performed the best in terms of transmission time (with minimal variation as well) despite being the worst in terms of percent error. Quality-based Djikstra, however, outpaced both the other two design options in terms of percent error and was better than bundle protocol (though not distance-based Djikstra) in terms of transmission time.

*Table 2* *Summary of the results from a single run of the Monte Carlo simulation, showing the mean, standard deviation, and standard error of the mean for the two metrics of interest.*

| Method | Percent error (%) | Transmission time (hrs) | | |
|---|---|---|---|---|
| | Mean | Mean | Standard deviation | Standard error |
| Bundle protocol | 57.6 | 2.703 | 0.994 | 0.0444 |
| Distance-based Djikstra | 77.8 | 1.1882 | 0.000129 | 5.759e-06 |
| Quality-based Djikstra | 36.4 | 1.926 | 0.970 | 0.0434 |

Figure 8, Figure 9, and Figure 10 below show the most frequent route taken by packets following bundle protocol, distance-based Djikstra, and signal quality-based Djikstra respectively. As is evident from the graphics, the MATLAB implementations of all three protocols behaved in the manner expected (see the protocol descriptions in Section 2 above), which provides confidence that they were implemented correctly.

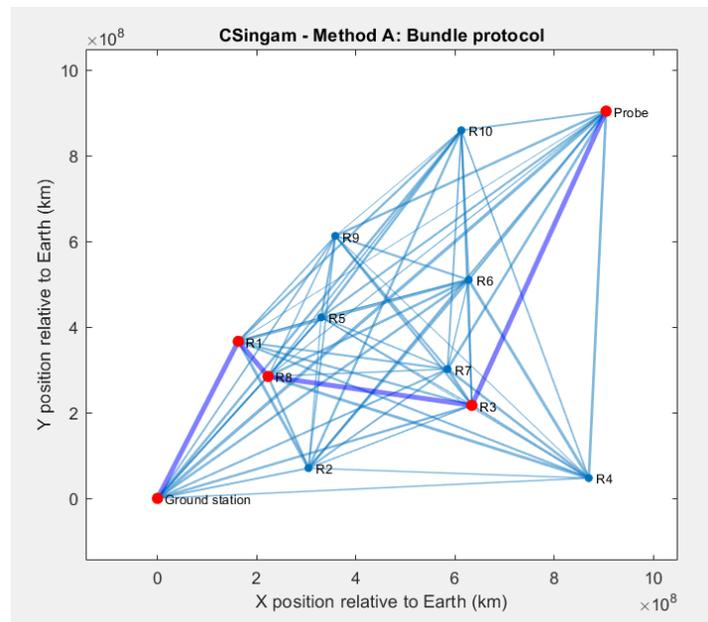

*Figure 8* *Most frequent path taken by packets using bundle protocol for the network associated with the results in Table 2.*

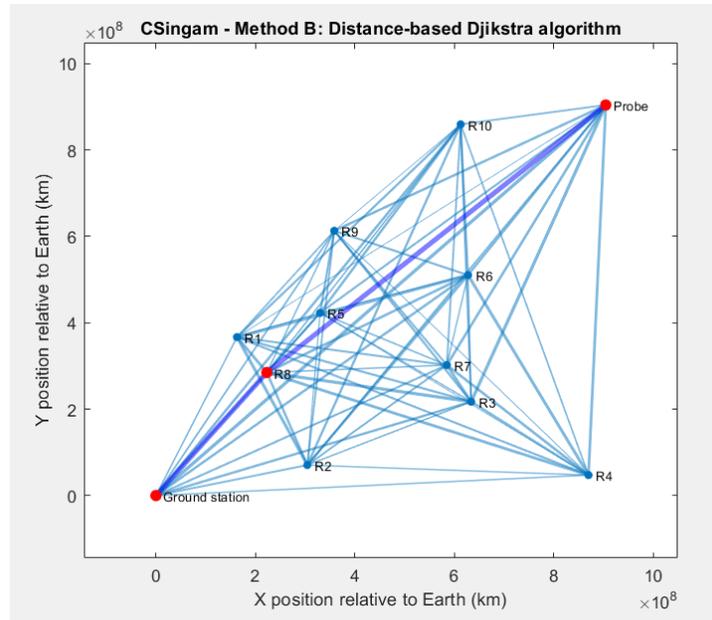

***Figure 9*** *Most frequent path taken by packets using distance-based Djikstra for the network associated with the Table 2 results.*

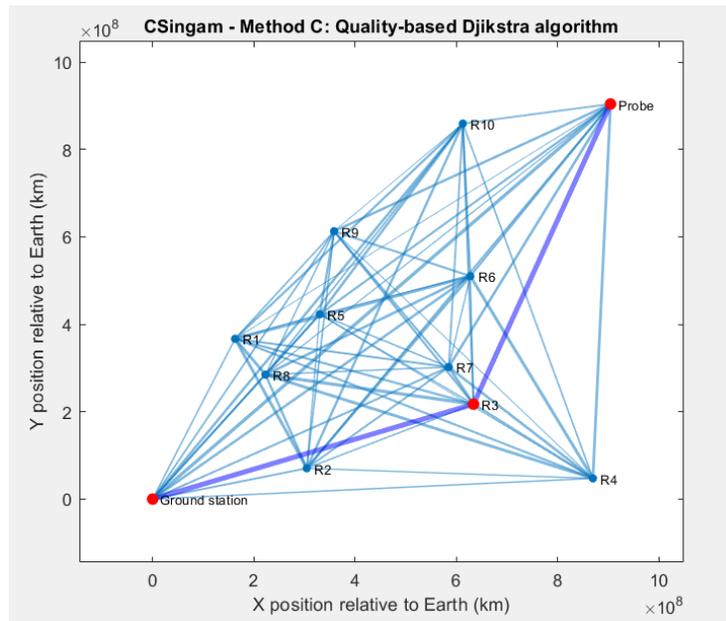

***Figure 10*** *Most frequent path taken by packets using quality-based Djikstra for the network associated with the Table 2 results.*

Since these are results for a single network, however, they do not reflect the variability in performance seen for each of the design options across different network configurations. Table 3 shows the performance of each of the three design options across five separate runs of the Monte Carlo simulation (and five different networks). The performance of the three routing protocols relative to each other remains consistent, with signal quality-based Djikstra showing the best results for percent error, distance-based Djikstra performing the best in transmission speed, and bundle protocol showing moderate results for both data integrity and the worst transmission time. Notably, distance-based Djikstra still performs worse than the current state-of-the-art methodology, bundle protocol, in terms of percent error. The other alternative protocol being evaluated, signal quality-based Djikstra, performs better than bundle protocol across both metrics.

*Table 3 Summary of the results from five runs of the Monte Carlo simulation, showing the mean, standard deviation, and standard error of the mean for the two metrics of interest.*

| Method | Percent error (%) | | | Transmission time (hrs) | | |
|---|---|---|---|---|---|---|
| | Mean | Standard deviation | Standard error | Mean | Standard deviation | Standard error |
| Bundle protocol | 56.440 | 7.410 | 3.314 | 3.120 | 0.684 | 0.306 |
| Distance-based Djikstra | 64.200 | 4.864 | 2.175 | 1.189 | 0.004 | 0.002 |
| Quality-based Djikstra | 41.040 | 4.498 | 2.011 | 1.820 | 0.280 | 0.125 |

Table 4 shows the results of the t-test, confirming that all of the differences observed in Table 3 between the different design options are in fact statistically significant (i.e. that choosing one option over another would result in a substantial difference in the metric) for both percent error and transmission time. Thus, performing a MAVF analysis is reasonable since it has been established that the choice of design option will matter and will cause a statistically significant difference in the metrics of interest.

*Table 4 Welch's t-test results comparing each of the design options with each of the other design options for both percent error and transmission time. Green cells indicate that the calculated p-value meets the standard for statistical significance (p<0.05).*

| Welch t-test results: percent error | | | | | Welch t-test results: transmission time | | | |
|---|---|---|---|---|---|---|---|---|
| vs | Bundle protocol | Distance-based Djikstra | Quality-based Djikstra | | Vs | Bundle protocol | Distance-based Djikstra | Quality-based Djikstra |
| Bundle protocol | - | 0.0914 | 0.0102 | | Bundle protocol | - | 0.00170 | 0.00500 |
| Distance-based Djikstra | 0.0914 | - | 8.758E-05 | | Distance-based Djikstra | 0.00170 | - | 0.00437 |
| Quality-based Djikstra | 0.0102 | 8.758E-05 | - | | Quality-based Djikstra | 0.00500 | 0.00437 | - |

Based on the results seen in Table 2 and Table 3, and taking the bundle protocol results as a benchmark for performance (what with it being the current preferred routing protocol for DTN contexts), it is almost unnecessary to perform a MAVF analysis since signal quality-based Djikstra outperforms bundle protocol across both metrics whereas the other non-SOA methodology, distance-based Djikstra, only outperforms bundle protocol in one metric (transmit time) and in fact performs significantly (as shown in Table 4) worse than baseline with regards to percent error. Nonetheless, the MAVF rankings are provided below in Table 5 to provide clear, unequivocal rankings of each design option.

*Table 5 MAVF results for each of the three design options (based on the metric means from the 5-run dataset).*

| MAVF analysis | | | |
|---|---|---|---|
| Value | | | |
| **Design option** | V(percent error) | V(transmit time) | MAVF |
| Bundle protocol | 0.33506 | 0 | 0.279217 |
| Distance Djikstra | 0 | 1 | 0.166667 |
| Quality Djikstra | 1 | 0.672774 | 0.945462 |

The MAVF results show quality-based Djikstra to be the best option, with a MAVF value that is over three times higher than the 2$^{nd}$ best option (bundle protocol). Distance-based Djikstra ranks the worst despite having the best value for transmit time.

## 6. Recommended course of action

The recommended course of action is to use signal-based Djikstra as the routing protocol for space-based DTN applications. In addition to being the best option to minimize transmission error, it also performs moderately well in terms of transmission time and outperforms the current standard for routing protocols, bundle protocol, across both metrics. The Welch's t-test results indicated that switching a network over from using either of the two other design options to using signal quality-based Djikstra would yield a statistically significant change in metrics, most notably an increase in data integrity on receipt (the more critical of the two parameters). Given that routing protocols are implemented in networks via software, and that software updates can be readily pushed remotely to satellites (by virtue of being communication instruments in and of themselves), it is feasible to implement a change in routing protocols without any hardware modifications to the system.

Since the MAVF results are meant to serve as a means of quantifying the relative practical value of the design options, it is also worth noting the methodologies which are not worth implementing (i.e. that performed worse than baseline). Distance-based Djikstra performed worse than baseline in a mission-critical metric (percent error), effectively eliminating any value the protocol's short transmission time might have had – after all, a rapid transmission has little utility if it risks compromising mission success significantly more than is considered standard.

Thus, it would be more accurate to assign a value of 0 to for the distance-based Djikstra methodology's transmit time; this would yield the corrected MAVF table seen below in Table 6.

*Table 6 The MAVF analysis results, with the values corrected for practicality.*

| MAVF analysis | | | |
|---|---|---|---|
| Value | | | |
| **Design option** | V(percent error) | V(transmit time) | MAVF |
| Bundle protocol | 0.33506 | 0 | 0.279217 |
| Distance Djikstra | 0 | 0 | 0 |
| Quality Djikstra | 1 | 0.672774 | 0.945462 |

### 7. Conclusion and lessons learned

In conclusion, signal quality-based Djikstra proved to be the best routing protocol for space-based DTN contexts, with bundle protocol (the current standard) coming in second and distance-based Djikstra perceived as a non-viable option.

It is also evident from the analysis results that the selection of a routing protocol for a generic space-based network is an important choice – for the three design options evaluated as part of this analysis, it was shown that a switch from any one of the design options to any other would result in statistically significant changes to both metrics of interest. Given this information, future work investigating the effect of these design options on other metrics may also be of interest.

Another possible avenue for future work is to include the effect of signal power and transmitter/receiver gain on the metrics of interest. For the sake of simplicity, this analysis was predicated on the assumption that all the satellites in the network had fixed hardware parameters (power and gain) since adjusting hardware parameters would impact perceived signal quality. While this assumption is acceptable when one is comparing simulated values with one another, a model that integrated hardware parameters and accurately represented signal dynamics would allow for the simulated outcomes of various routing protocols to be faithful representations of what might otherwise be seen with real-world implementations of such protocols – thereby enabling the comparison of simulated values to the quantitative benchmarks (i.e. threshold values) that would be used to evaluate the performance of a real-world system based on the metrics of interest.

It is recommended, however, that any future work be pursued using a Python implementation of the models/simulations rather than a MATLAB implementation. The MATLAB implementation proved to be resource intensive; the author has observed that the built-in MATLAB functions utilized for instantiating digraphs, performing shortest-path analyses, etc. use numerous function calls rather than implementing various routines natively in the script which likely is the reason why running a relatively short (< 300 lines) piece of code proved so taxing to the author's computer system. Python, however, uses more self-contained implementations of various methods since most of its libraries are open-source entities created by members of the public and are thus less likely to have numerous dependencies that slow down processing. The current implementation of the Monte Carlo simulation had an average of 302.308 +/- 13.228 seconds when run in MATLAB; it is anticipated that a Python implementation, if done correctly, would have a noticeably shorter runtime.